\documentclass[epj]{svjour}
\usepackage{latexsym}
\usepackage{graphics}
\DeclareTextSymbol{\degree}{OT1}{23}

\begin{document}

\title{Probing the micromechanics of a multi-contact interface at the onset of frictional sliding}
\author{A. Prevost\inst{1}  \thanks{\email{alexis.prevost@lps.ens.fr}} \and J. Scheibert\inst{2} \and G. Debr\'egeas\inst{1}
}                     
%
%
\institute{Laboratoire Jean Perrin, UPMC Universit\'e Paris 6, CNRS FRE 3231, Ecole Normale Sup\'erieure, 24 rue Lhomond, 75005 Paris, France \and Laboratoire de Tribologie et Dynamique des Syst\`emes, CNRS, Ecole Centrale de Lyon, Ecully, France}
\date{Received: date / Revised version: date}
%
\abstract{
Digital Image Correlation is used to study the micromechanics of a multi-contact interface formed between a rough elastomer and a smooth glass surface. The in-plane elastomer deformation is monitored during the incipient sliding regime, \textit{i.e.} the transition between static and sliding contact. As the shear load is increased, an annular slip region, in coexistence with a central stick region, is found to progressively invade the contact. From the interfacial displacement field, the tangential stress field can be further computed using a numerical inversion procedure. These local mechanical measurements are found to be correctly captured by Cattaneo and Mindlin (CM)'s model. However, close comparison reveals significant discrepancies in both the displacements and stress fields that reflect the oversimplifying hypothesis underlying CM's scenario. In particular, our optical measurements allow us to exhibit an elasto-plastic like friction constitutive equation that differs from the rigid-plastic behavior assumed in CM's model. This local constitutive law, which involves a roughness-related length scale, is consistent with the model of Bureau \textit{et al.} [Proc. R. Soc. London A \textbf{459}, 2787 (2003)] derived for homogeneously loaded macroscopic multi-contact interfaces, thus extending its validity to mesoscopic scales.}

\maketitle

\section{Introduction}
\label{intro}

The transition from static to sliding friction is a crucial process in various fields, ranging from contact mechanics \cite{JohnsonBook}, earthquakes dynamics \cite{ScholzBook} to human/humanoid object grasping \cite{Johansson2009}. In the classical Amontons-Coulomb's framework, when two solids are brought in contact under a normal load $P$ and subjected to a shear force $Q$, no relative motion occurs until $Q$ exceeds some threshold value $Q_s = \mu_s P$ where $\mu_s$ is called the static friction coefficient. However, in most real situations, the transition from static to dynamic friction does not follow this ideal simple scenario. As soon as $Q>0$, partial slippage generally sets in owing to the large stress heterogeneity within the contact zone, which depend on the geometry of the objects in contact as well as on the loading conditions. Understanding this incipient sliding regime thus requires to gain access to the interfacial micromechanics within the contact zone.

In the past ten years, several experimental groups developed new optical methods to obtain spatially resolved mechanical measurements  \cite{Baumberger2002,Rosakis2004,Rubinstein2009,FinebergScience2010,Maegawa2010}, which triggered intense subsequent theoretical and numerical investigations \cite{Braun2009,Scheibert2010,DiBartolomeo2010,Scheibert2011,Bouchbinder2011,Amundsen2012,Kammer2012}. Fineberg and collaborators studied the onset of sliding of a multi-contact interface, in a plane-plane contact configuration, submitted to an adiabatic tangential loading \cite{Rubinstein2009,FinebergScience2010}. Using fast imaging of the interface illuminated with an evanescent laser sheet, they were able to measure local changes in the real area of contact. This simple optical measurement allowed them to reveal that, prior to macroscopic sliding, series of dynamical rupture fronts traveled along the interface. In their experiments however, the contact was one-dimensional. Chateauminois and collaborators have considered more realistic, fully two-dimensional, contacts \cite{Chateauminois2008,Chateauminois2010}. By patterning a smooth elastomer's surface with a regular grid of micro-markers, they were able to monitor, using Particle Image Velocimetry techniques, the entire 2D displacement field at the interface. They applied this procedure to a smooth sphere loaded against a \textit{smooth} flat elastomer block in a \textit{torsional} configuration \cite{Chateauminois2010}. The authors showed that the transition from static to kinetic friction involves an annular micro-slip front propagating from the outer edge of the circular contact towards the center. Using an inversion procedure, they computed the interfacial shear stress field from the measured displacements field. We emphasize that in their experiments, macroscopic adhesion was important due to the smoothness of the surfaces in contact. In most practical situations however, interfaces are rough at microscopic length scales, giving rise to a multi-contact frictional interface at which macroscopic adhesive effects are strongly reduced. 

In this paper, we expand on this latter two-dimensional approach. We report on micromechanical measurements at the onset of sliding between a smooth sphere and a flat elastomer block whose surface is microscopically rough. There are several important differences with respect to \cite{Chateauminois2010}. First, the interface is of the multi-contact type. Second, the loading is linear instead of torsional. Third, the displacement field measurement is truly non-invasive as we take advantage of the optically diffusive nature of the multi-contact interface to extract the displacement fields at all times using Digital Image Correlation (DIC), from which the interfacial stress fields can be further computed.

In Section \ref{sec:1}, we present the experimental setup and samples preparation and show how the displacement field is extracted from successive images of the interface. Macroscopic force measurements and resulting displacement fields are then presented in Section \ref{sec:2}. Section \ref{sec:3} details the procedure used to compute the interfacial stress fields from the measured displacement fields. In Section \ref{sec:4}, these results are compared to the predictions of Cattaneo and Mindlin (CM)'s classical contact mechanics model. In Section \ref{sec:5}, we interpret the observed deviations as the effect of the rough layer, which introduces a micrometric length scale in the problem. Eventually, in Section \ref{sec:6}, general conclusions and perspectives are drawn.

\section{Experiment, materials and methods}
\label{sec:1}

\subsection{Experimental setup}
\label{subsec:1}

\begin{figure}
\resizebox{1.0\hsize}{!}{\includegraphics*{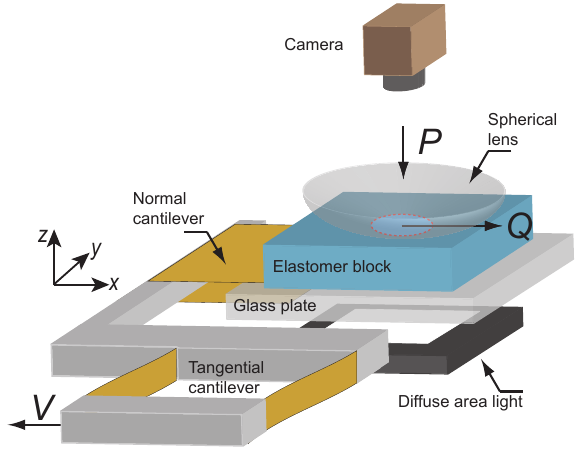}}
\caption{(Color online) Sketch of the experimental setup. The normal cantilever used to measure $P$ has a stiffness $k_N=689 \pm 5$ N.m$^{-1}$, while the tangential cantilever used to measure $Q$ has a stiffness $K_T=9579 \pm 25$ N.m$^{-1}$.} 
\label{Fig1}
\end{figure}

Figure~\ref{Fig1} shows a sketch of the experimental setup. It consists of a planoconvex glass lens (optical grade, Thorlabs LA1301, BK7, radius of curvature $R=128.8$ mm) glued onto a lens holder and rigidly attached to an optical table. The lens surface is in frictional contact against a thick elastomer block maintained by Van der Waals adhesion to a supporting glass plate.  The latter is connected \textit{via} a set of two orthogonal cantilevers to a translation stage that can be driven at constant velocity $v$ using a DC actuator (LTA-HL, Newport Inc.). Two capacitive position sensors (respectively MCC-20 and MCC-10, Fogale Nanotech), each facing the mobile part of one cantilever, allow one to measure both $P$ and $Q$, respectively the normal and tangential (shear) force, with a 1 mN resolution in the range [0--2 N]. Both force signals are digitized and recorded at a sampling rate of 1 kHz using a NI-PCI6251 DAQ board. Imaging of the contact is done by illuminating the interface through the transparent elastomer block with a white LED diffusive array and a long-working distance Navitar objective. Images of the interface are recorded with a CCD camera (Redlake ES2020M, $1600\times1200$ pixels$^2$, 8 bits, 24 frames/s at maximum). Synchronization between the camera and the DAQ acquisition device is ensured by having the DAQ board trigger the camera.   
 
\subsection{Sample preparation and characterization}
\label{subsec:2}

The elastomer block ($50 \times 50$ mm, thickness $h=15$ mm) is made of crosslinked PolyDimethylSiloxane (PDMS Sylgard 184, Dow Corning). It is obtained by mixing in a 10:1 stoichiometric ratio a PDMS melt and a cross-linker agent in a rectangular mold. The mixture is cured for 48 hours in an oven at $70\degree$C. The free surface of the elastomer block is rendered rough by mechanically abrading the lid's upper surface with a Silicon Carbide powder solution of typical grain size 17~$\mu$m. After careful demoulding, the surface roughness of the PDMS sample is characterized with an optical profilometer (M3D, Fogale Nanotech). Its height power spectrum is found to decay as a power law from a maximum value of $\sim 30 \mu$m down to the micrometer scale. The characteristic thickness $h$ of the rough interface, defined as the standard deviation of the height distribution, is measured to be $0.595 \pm 0.013 \, \mu$m. 

Such PDMS elastomers have been reported to have a bulk elastic Young's modulus $E$ in the range [2--4 MPa] \cite{Chateauminois2008,ScheibertEPL2008,ScheibertJMPS2009,LorenzPersson2009} (depending on the preparation protocol) and a Poisson's ratio $\nu$ close to 0.5 \cite{PoissonRatio}. For the sample used in these experiments, a JKR test is performed between its smooth back side and the bare glass lens, yielding a Young's modulus $E=3.43 \pm 0.05$~MPa. For the friction experiments, the glass lens surface is passivated in a PerfluoroDecylTricloroSilane saturated vapor phase in order to reduce the macroscopic PDMS/glass adhesion and to minimize heterogeneities in the interfacial properties. Prior to each experiment, both glass and PDMS surfaces are cleaned with ethanol and dried with filtered air.

\subsection{Contact imaging and displacement fields measurements}
\label{subsec:3}

\begin{figure}
\resizebox{1.0\hsize}{!}{\includegraphics*{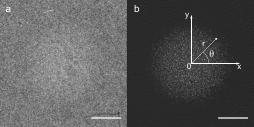}}
\caption{\label{Fig2} (a) Image of the interface at $P=0.1$ N. (b) Difference between image (a) and a reference image without contact. For both images, the white bar is 1 mm long. On (b) are shown the $x$ and $y$ axis along with the $r,\theta$ coordinates system.}
\end{figure}

\subsubsection{Contact imaging}
\label{subsubsec:3:1}
Figure~\ref{Fig2}a shows a typical image of the elastomer/glass interface under normal load. This image is obtained in transmission geometry by  illuminating the sample with a diffuse white light. The interface appears spatially heterogeneous as a result of the diffusive nature of the rough layer. In the contact region, additional bright spots are present corresponding to the micro-junctions that favorably transmit light at the glass-PDMS interface. At the chosen magnification, images have a field of view of about $11.2 \times 8.4\,$mm (with a pixel size of $\sim 7.04~\mu$m). The contact region is difficult to visualize from the raw image (Fig.~\ref{Fig2}a) but can be clearly identified in Fig.~\ref{Fig2}b by subtracting a reference background image recorded prior to the contact\footnote{Careful image difference was done to take into account a possible rigid-body displacement of the PDMS block between the unloaded (reference) and loaded situations.}. This operation allows one to reveal hundreds of contact-induced micro-junctions contained within a circular (apparent) contact region (Fig.~\ref{Fig2}b for $P=0.1$~N). These characteristic features of the frictional joint remain true for the explored range of normal load $P$ used in these experiments. 

\begin{figure}
\resizebox{1.0\hsize}{!}{\includegraphics*{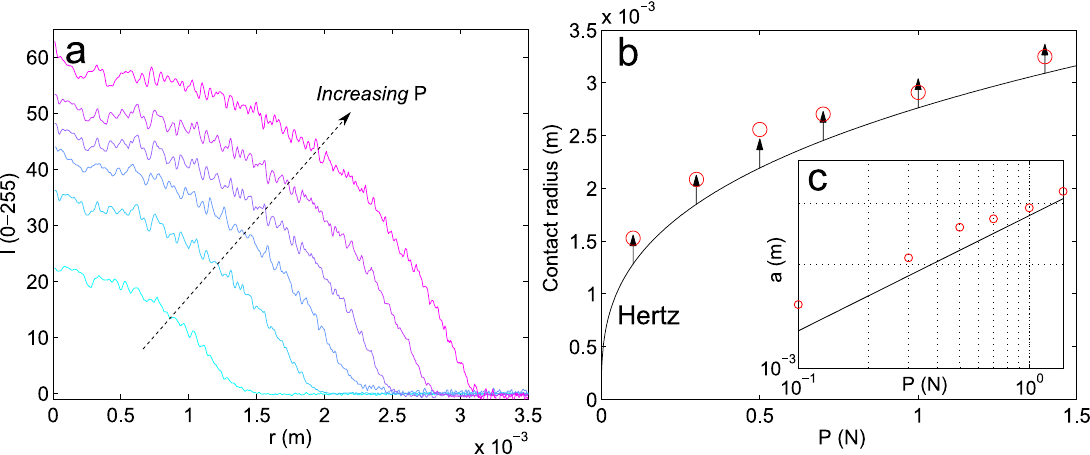}}
\caption{\label{Fig3} (Color online) (a) Angularly averaged intensity profiles $I(r)$ for increasing loads $P$, with $r=0$ located at the center of the apparent contact. The lowest $P$ curve is in cyan, while the highest one is in pink. This color code is kept within this paper to describe any normal dependence of the variables involved. (b) Apparent contact radius versus $P$. The red circles correspond to the measured apparent contact radius $a$ obtained by taking the minimal value of $r$ for which $I(r)=0$. The thin black line is Hertz contact radius $a_H$ with $E=3.43$~MPa. Black vertical arrows have the same length, equal to $\sqrt{Rh}$. (c) Inset: Same as (b) on a log-log scale. The solid black line corresponding to Hertz prediction has a slope of 1/3.}
\end{figure}

Taking advantage of the axial symmetry of the contact, the apparent contact radius is computed in the following way. Centers of the apparent contact $r=0$ are first determined using a standard center of symmetry search algorithm. For each load $P$, the image intensity $I(x,y)$ is then averaged azimuthally over the angle $\theta$ (Fig.~\ref{Fig3}a) \footnote{In order to keep the statistical errors constant, the number of pixels within each annulus is kept constant, equal to 3000. Thus, the width of the annuli decrease with their radius $r$.}. The apparent radius of contact $a$ is then estimated as the minimum radius such that $I(r)=0$. Figure~\ref{Fig3}b shows the resulting contact radius $a$ and its load dependence along with Hertz \footnote{Hertz contact is the contact between two non-conforming elastic half-planes having smooth surfaces and well defined radii of curvature. For instance, the Hertz contact radius $a_H$ in the case of a rigid sphere pressed against an elastic plane is $a_H=\left(\frac{3PR(1-\nu^2)}{4E}\right)^{1/3}$, with $P$ the normal load, $R$ the sphere radius, and $E$ and $\nu$ the Young's modulus and Poisson's ratio of the elastic plane.} contact radius $a_H$ derived for $E=3.43$~MPa. As clearly seen, the estimated radius $a$ is systematically larger than $a_H$. This deviation can be accounted for by taking into account the multi-contact nature of the interface. As established by Greenwood and Tripp \cite{GreenwoodTripp1967,PhD-JS-2008}, the surface roughness extends the apparent contact region by a quantity of the order of $\sqrt{Rh}~\sim$~280~$\mu$m with respect to the smooth configuration. As shown on Fig.~\ref{Fig3}b, such a correction to Hertz's contact radii does allow one to recover the measured apparent contact radii $a$. 

\subsubsection{Displacement field measurements}

\begin{figure}
\resizebox{1.0\hsize}{!}{\includegraphics*{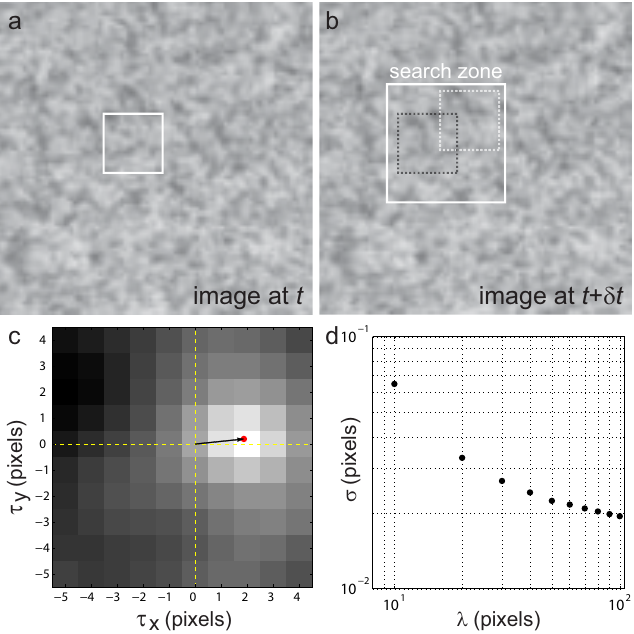}}
\caption{\label{Fig4} (Color online) Principles of the DIC measurement. (a) Image at time $t$. (b) Image at time $t + \delta t$. (c) 2D correlation function $C(\tau_x,\tau_y)$ obtained by correlating the white line delimited box in (a) (size $\lambda = 20$ pixels) with boxes in (b) (dashed lines), displaced by $(\tau_x, \tau_y)$ within the search zone. The red dot indicates the sub-pixel location of the maximum of the correlation function. The black arrow shows the resulting displacement vector. (d) Log-log plot of the standard deviation $\sigma$ of the displacements distribution versus the box size $\lambda$.}
\end{figure}

Interfacial displacement fields were computed from snapshots acquired at 8 frames/s ($\Delta t=0.125$~s), and 4 frames/s for the highest $P$ ($\Delta t=0.25$~s), using a Digital Image Correlation technique (DIC) (see \textit{e.g.} \cite{HildRoux2006} and references therein). This method consists in finding, for a given sub-image centered at position ($x,y$) in a reference frame, the displacement ($u_x,u_y$) that provides the maximum intensity correlation with a subsequent (deformed) image (Figs.~\ref{Fig4}a and b). A 2D correlation function (Fig.~\ref{Fig4}c) was computed using a direct calculation. Sub-pixel resolution was achieved by fitting the correlation function with a 2D Gaussian surface using the pixel of maximum correlation and its 8 nearest neighbors. The error on the displacements was evaluated using a series of images of the surface of the elastomer block, not in contact and uniformly displaced at constant velocity along the $x$ direction. DIC was performed between an image at $t=0$ and images at increasing times $t$, allowing one to extract the displacements $u_x$ between $\sim 0.14$ pixels and $\sim 14$ pixels. The error was then taken as the standard deviation of the $u_x-vt$ displacements distribution. This error was found to decrease with the box size $\lambda$, in the range [10--100]~pixels (Fig.~\ref{Fig4}d). The optimal $\lambda$ was chosen based on the best compromise between spatial resolution and displacement measurement accuracy. For the current experiments, the smallest apparent contact radius is $\sim 217$~pixels long. In order to extract at least 10 independent displacement measurements along the contact, the displacement fields were computed with $\lambda = 20$ pixels, on a $10$ pixels wide square grid, yielding a resolution on the displacement of $\sim 0.033$ pixel ($\sim 232$ nm).

\section{Force and displacement fields}
\label{sec:2}

We performed a series of 6 experiments in which the elastomer block was driven at a prescribed velocity\footnote{In practice, the actual velocity was slightly smaller than the prescribed velocity, decreasing monotonically from $4.9 \mu$m/s at $P=0.1$~N to $4.76 \mu$m/s at $P=1.4$~N.} $v=5~\mu$m/s, under constant normal force $P$ in the range [0.1--1.4~N]. For all runs, $P$ was found to vary by less than 1\% over the duration of the experiment and the apparent contact zone remains circular. The velocity $v$ was chosen low enough for visco-elastic interfacial dissipation to be negligible \cite{Ronsin-ProcRSocLondA-2001}. Since the normal loading of the contact produces a significant shear force due to the small coupling between normal and lateral motion of each cantilever, the contact was manually renewed prior to each experiment, until the initial shear force $Q$ was less than 1\% of $P$. During this separation procedure, no measurable pull-off force was observed, indicating that adhesion forces are negligible \cite{Shull-MatSciEngR-2002,Fuller-Tabor-ProcRSocLondA-1975} at the contact length scale.

\begin{figure}
\resizebox{1.0\hsize}{!}{\includegraphics*{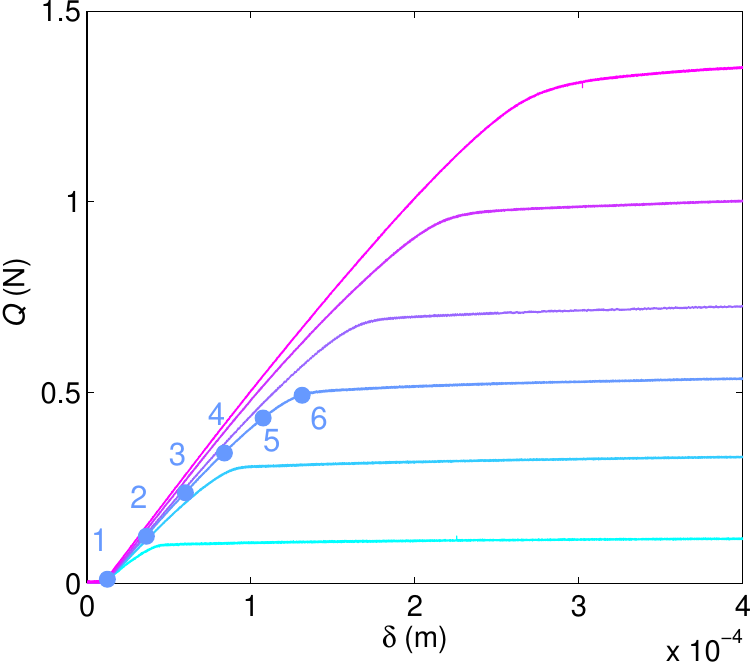}}
\caption{\label{Fig5} (Color online) Shear force $Q$ versus $\delta=v t$ for loads $P=0.1, 0.3, 0.5, 0.7, 1.0, 1.4$ N (bottom to top). The color code is the same as in Fig.~\ref{Fig3}a. Numbered disks from 1 to 6 correspond to the 6 instants at which the displacement fields are displayed in Figs.~\ref{Fig6} and \ref{Fig10}.}
\end{figure}

Figure~\ref{Fig5} shows the measured evolution of the tangential $Q$ versus $\delta=v t$ for the 6 values of the normal load $P = \{0.1, 0.3, 0.5, 0.7, 1.0, 1.4\}$~N. Each curve exhibits two distinct phases: an initial quasi-linear loading associated with the incipient sliding regime, followed by a plateau associated with a macroscopic steady sliding regime. The transition between these two regimes involves a monotonous decrease of the slope. In particular, we do not observe any static friction peak in the loading curve, which hampers a direct determination of the force threshold $Q_s$ at which macroscopic sliding sets on from the sole global force measurements. $Q_s$ was thus determined using the measured displacement fields at the center of the contact as detailed further down.

\begin{figure*}
\resizebox{1.0\hsize}{!}{\includegraphics*{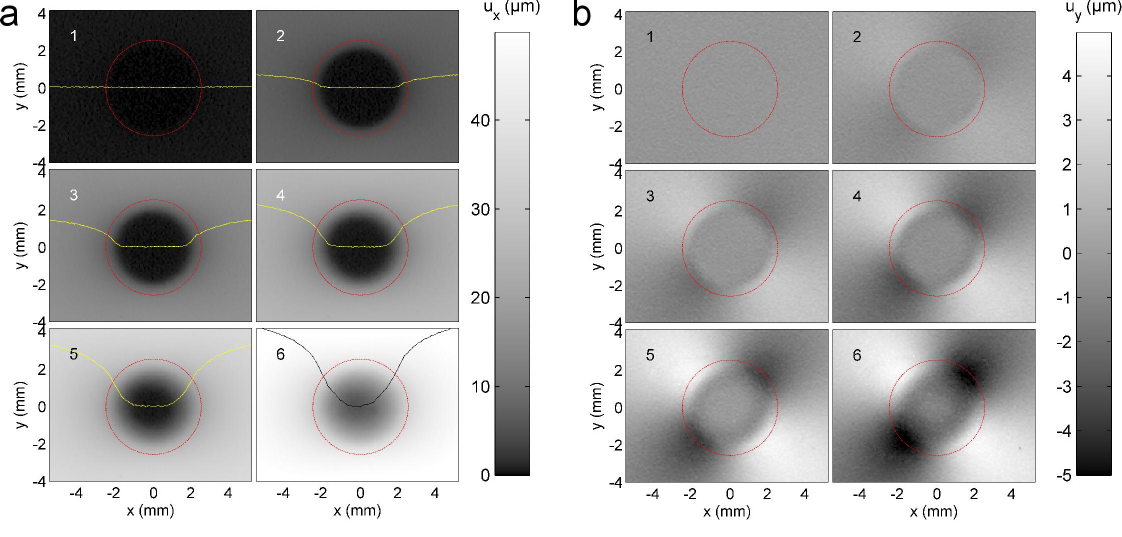}}
\caption{\label{Fig6} (Color online) Snapshots of the 2D displacement fields $u_x$ and $u_y$ at $P=0.5$~N for the loading experiment of Fig.~\ref{Fig5} taken at instants labeled 1 to 6. On all displacement fields, the red dashed circle delimits the apparent contact area of radius $a$. The yellow curves (respectively black curve) on snapshots indexed 1 to 5 (respectively 6) are cuts of the $u_x$ 2D displacements fields at $y=0$ and are meant as visual guides. (a) $u_x$ displacement fields. (b) $u_y$ displacement fields.}
\end{figure*}

Displacement fields $u_x$ and $u_y$ at time $t$ were determined by correlating images at time $t$ with a reference image at $t=0$ prior to any tangential loading using the algorithm described earlier. Since the lens holder is not infinitely rigid, any applied shear force is expected to induce minute rigid body displacement of the lens that needs to be subtracted to the measured displacement $u_x$. To quantify this compliance effect, we independently measured the displacement of the lens holder while tangentially loading the elastomer block under controlled shear force using the same correlation technique. The solid lens displacement was found to vary linearly with $Q$, yielding a shear  stiffness $\approx 0.68~\mu$m~N$^{-1}$. Actual interfacial displacements along the $x$ direction were then corrected for this finite compliance effect. 

\begin{figure}
\resizebox{1.0\hsize}{!}{\includegraphics*{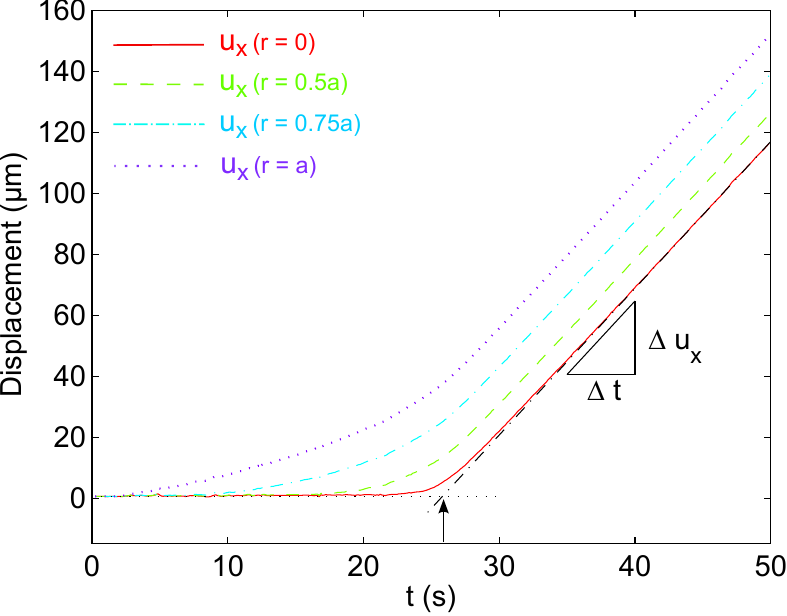}}
\caption{\label{Fig7} (Color online) Angularly averaged radial displacement $u_x(r)$, with $r=0$ being the center of the contact, for values of $r=0, 0.5a, 0.75a, a$ at a normal load $P=0.5$~N. The ratio $\frac{\Delta u_x}{\Delta t}$ is $\sim 4.81 \mu$m/s. The horizontal black dotted line is the $y=0$ axis. The dashed dotted line is a linear fit of $u_x(r=0)$ at long times. The intersection of both lines, shown with the vertical black arrow, was arbitrarily chosen as the definition for the onset time of macroscopic sliding.}
\end{figure}

\begin{figure}
\resizebox{1.0\hsize}{!}{\includegraphics*{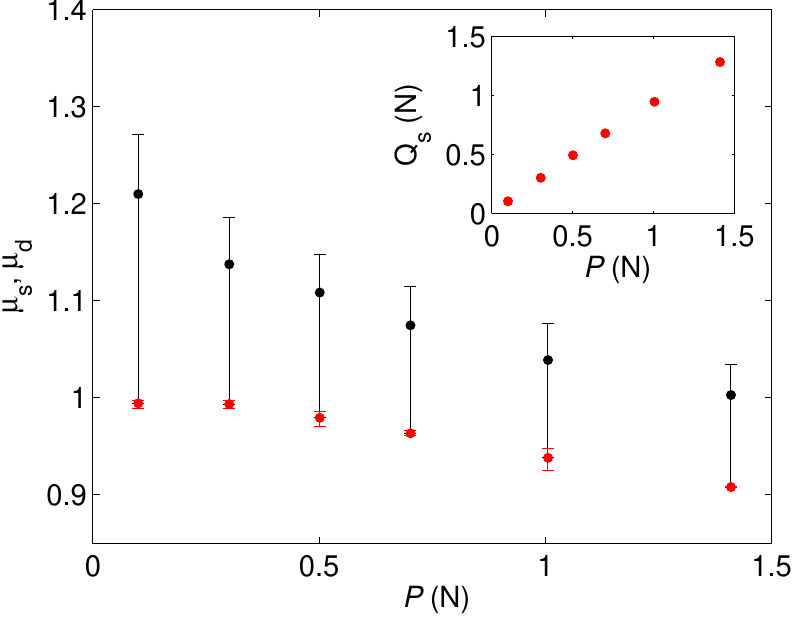}}
\caption{\label{Fig8} (Color online) Friction coefficients $\mu_s$ (lower red disks) and $\mu_d$ (upper black disks) versus $P$. $\mu_s$ is defined as the ratio $Q_s/P$ where $Q_s$ (shown in the inset) has been determined as described earlier and schematically shown on Fig.~\ref{Fig7}. $\mu_d$ is taken as the mean of $Q/P$ for all $Q \geq Q_{s}$. Error bars are deduced from its minimum and maximum values over this range.}
\end{figure}

Figure~\ref{Fig6} shows the typical $u_x$ and $u_y$ displacement fields at $P=0.5$~N for all 6 positions on Fig.~\ref{Fig5}. As soon as $Q$ increases, the displacement in the outer region of the apparent contact increases. In contrast, the measured displacement remains essentially null within a central circular region. As $Q$ increases, the radius $c$ of this stick region decreases and eventually vanishes, marking the onset of the macroscopic sliding phase.
Figure~\ref{Fig7} shows the time-evolution of the displacement $u_x(r)$, averaged over the azimuthal angle $\theta$, for different radii $r=\{0, 0.5a, 0.75a, a\}$ at a normal load $P=0.5$~N. In the fully developed sliding regime, the displacement $u_x$ varies linearly and uniformly with time $t$ (while $u_y$ remains stationary), indicating that the PDMS sample is sliding as a whole at the driving velocity. In the incipient sliding regime, the time evolution of the displacement $u_x$ strongly depends on $r$. As previously mentioned, the transition to sliding initiates in the outmost part of the contact zone and is completed when the center region (corresponding to the $r=0$ curve) starts sliding. In order to properly extract the onset time $t_s$ of macroscopic slippage, the central displacement time signal $u_x(r=0)(t)$  is used. The time $t_s$ is evaluated as the intersection between the linear fit of $u_x(r=0)(t)$ at long times and the time axis. This time of macroscopic slippage allows us to unambiguously identify  a threshold force $Q_s$ and an associated macroscopic friction coefficient $\mu_s=Q_s/P$. The dynamical friction coefficient is further defined as $\mu_d = Q(t \ge t_s)/P$. Figure~\ref{Fig8}, which shows both friction coefficients versus $P$, indicate that $\mu_s$ and $\mu_d$ both display a slow decay with $P$, as it is usually found for elastomers \cite{ScheibertJMPS2009,WandersmanPRL2011}.

\section{Tangential stress fields}
\label{sec:3}

Tangential stress fields at the contacting interface were derived using a Green's tensor inversion procedure as described in \cite{Chateauminois2008}. For a semi-infinite elastic medium, the Green's tensor characterizes the displacements at the interface induced by a point force applied at the free surface \cite{LandauLifshitz}. In the limit of a semi-infinite and incompressible elastic body, two assumptions which are well suited to our experiments, the lateral and vertical displacements are decoupled, allowing one to express the lateral displacements as a function of the lateral shear stresses only \cite{ScheibertJMPS2009}. For a point loading ($Q_x, Q_y$), the surface displacements $u_x$ and $u_y$ are thus given respectively by 

\begin{eqnarray}
	u_x = G_{xx} Q_x + G_{xy} Q_y \nonumber\\
  u_y = G_{yx} Q_x + G_{yy} Q_y
\end{eqnarray}
with the components of the Green's tensor $\underline{G}$ given by 

\begin{eqnarray}
  G_{xx} = \frac{3}{4 \pi E} \left(\frac{1}{r}+\frac{x^2}{r^3}\right) \nonumber\\
  G_{xy} = G_{yx} = \frac{3}{4 \pi E} \left(\frac{xy}{r^3}\right) \nonumber\\
  G_{yy} = \frac{3}{4 \pi E} \left(\frac{1}{r}+\frac{y^2}{r^3}\right)
  \label{EqnG}
\end{eqnarray}

For an extended contact, $u_x$ and $u_y$ are obtained by convolving $\underline{G}$ with the shear stress at the interface $\underline{\sigma}$ and can be formally written as 

\begin{equation}
  u_{i} = G_{ij} * \sigma_{jz}
  \label{EqnConvol} 
\end{equation}
where subscripts $i,j$ stand for $x$ or $y$. The stress fields $\sigma_{xz}$ and $\sigma_{yz}$ can then be obtained by deconvolution. This was done as in \cite{Chateauminois2008} using a classic iterative Van-Cittert algorithm. The stress at step $n+1$ is obtained by adding to the one at time $n$ a corrective term\footnote{In practice, an initial guess at step $n=0$ is taken as proportional to the experimental displacements. Such an iterative deconvolution procedure can very rapidly lead to numerical divergence if not slowed down. Empirically, only 5 percent of the difference in displacements is thus added to the stress at step $n$, enabling convergence in typically hundreds of iterations.} proportional to the difference between the experimental displacement and the convolved one obtained using Eq. \ref{EqnConvol}. Convergence was considered to be attained when the {\it rms} difference between the calculated and the measured displacements was less than the displacement resolution, i.e. 0.033 pixels.

\begin{figure}
\resizebox{1.0\hsize}{!}{\includegraphics*{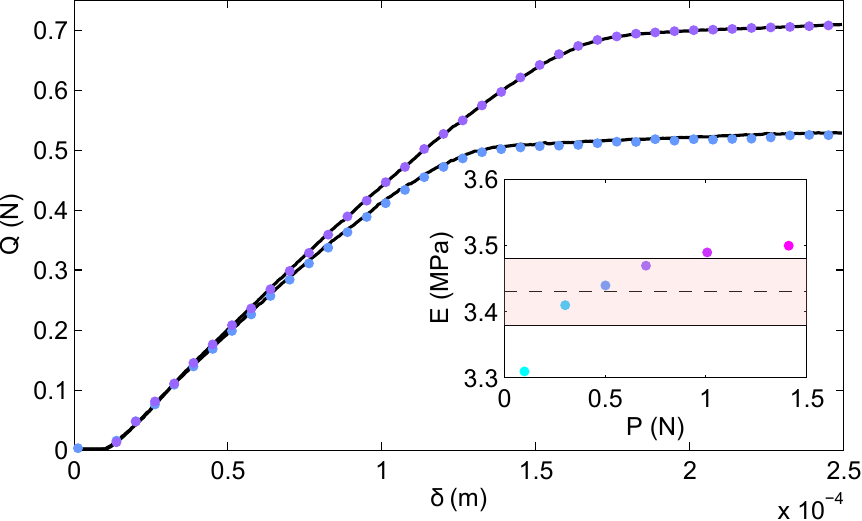}}
\caption{\label{Fig9} (Color online) Comparison between $Q$ versus $\delta$ as measured in the experiment (black solid lines) and obtained by summing up the shear stress within the contact zone, at $P=0.5$ N and $P=0.7$ N. For clarity, 1 experimental point out of 10 is shown. Inset: Optimum $E$ values versus $P$. The shaded area represents the value $E=3.43 \pm 0.05$~MPa obtained with the JKR test (\textit{see} Section \ref{subsec:2}). }
\end{figure}

The inversion procedure provides shear stress fields in units of the Young's modulus $E$ (\textit{see} Eqs. \ref{EqnG} and \ref{EqnConvol}). The value of $E$ can then be directly fitted from the comparison between the shear force signal $Q(t)$, obtained through spatial integration of the calculated shear stress over the contact, and the actual measured force signal. As shown in Fig.~\ref{Fig9} for $P=0.5$~N and $P=0.7$~N, the match between both signals is very satisfactory, which validates the inversion method. The extracted Young modulus shows a weak dependence with the load (\textit{see} inset) for which we have no clear explanation. However, except for the two extreme load values, it remains within the error bars of $E$ as independently measured with the JKR test. Note that in the determination of the stress fields, we have used the fitted values for $E$ rather than the JKR value.

\begin{figure}
\resizebox{1.0\hsize}{!}{\includegraphics*{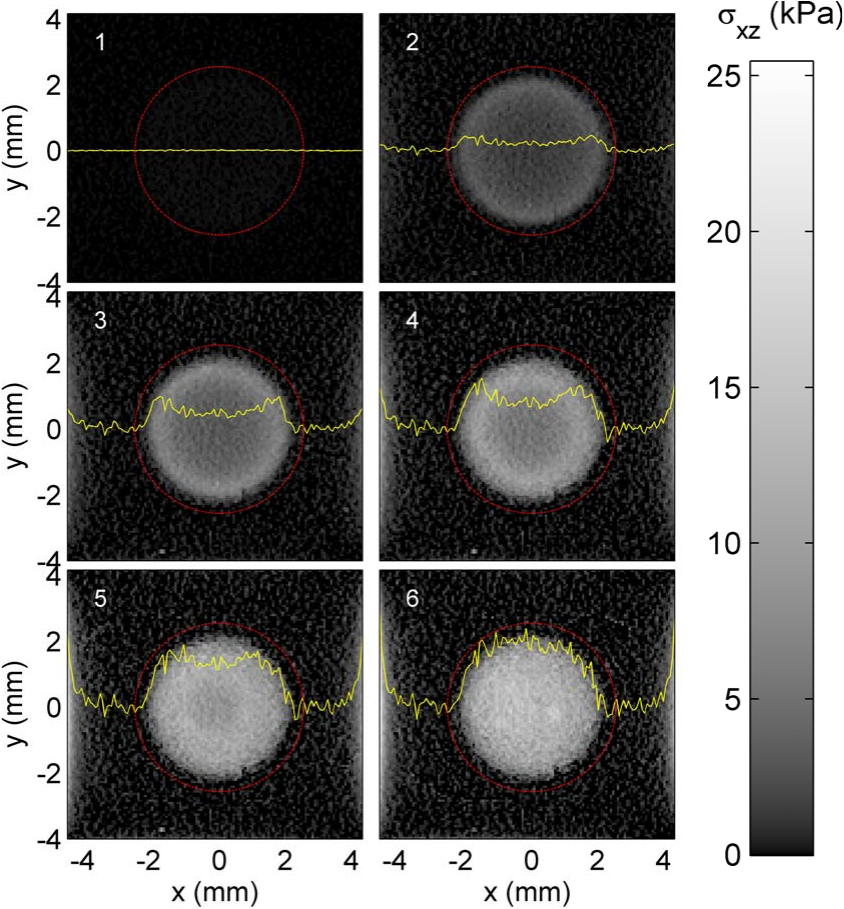}}
\caption{\label{Fig10} (Color online) Snapshots of the 2D stress fields $\sigma_{xz}(x,y)$ at $P=0.5$~N for the loading experiment of Fig.~\ref{Fig5} taken at instants labeled 1 to 6. The yellow curves are cuts along $y=0$ and are intended to ease the visualization of a lower stress region around the center. On all displacement fields, the red dashed circle delimits the apparent contact area.}
\end{figure}

Figure~\ref{Fig10} displays the shear stress fields $\sigma_{xz}$ computed at successive moments (indicated by the 6 labeled dots in Fig.~\ref{Fig5}) during the incipient sliding phase. As expected, in all 6 configurations, the shear stress vanishes outside the apparent contact zone whose border is indicated by a red dashed circle. Within the contact, $\sigma_{xz}$ is radially symmetric with respect to the center of the contact. Once macroscopic sliding has initiated (position 6), $\sigma_{xz}$ is maximum at the center of the contact and decreases continuously towards the edge of the contact. During the transient loading however, the shear stress exhibits a local minimum at the center of the contact. 

\section{Comparison with models predictions}
\label{sec:4}

These precise local mechanical measurements are directly amenable to comparison with existing theoretical models of incipient sliding. The first model, for a non-adhesive elastic sphere-on-plane contact, was derived independently by Cattaneo and Mindlin (CM) \cite{Cattaneo-RANL-1938,Mindlin-JAppMech-1949}. Since then, this classic model has been refined and extended in various ways (see for instance \cite{JohnsonBook,HillsBook,Ciavarella1998} and references therein), for instance by introducing macroscopic adhesion \cite{SavkoorThesis,SavkoorBriggs1977} or the elasto-plasticity of the materials \cite{Etsion2010}. Since (i) adhesive forces were found to be negligible in our experiments (no measurable pull-off force upon retraction of the contact) and (ii) PDMS was used far from its plastic limit, our measurements were compared to CM's model. 

CM's calculations assume that (1) both surfaces are \textit{smooth}, (2) the pressure distribution $\sigma_{zz}$ within the contact is unchanged upon shearing and given by Hertz contact theory, and (3) Amontons-Coulomb's law of friction is valid locally at any position within the contact, \textit{i.e.} slip occurs wherever the shear stress $\sigma_{xz}$ reaches $\mu \sigma_{zz}$, $\mu$ being the macroscopic friction coefficient\footnote{CM's model postulates the existence of a single, stress-independent friction coefficient, \textit{i.e.} static and dynamical friction coefficients are equal, and thus predicts no static overshoot in the $Q$ curve.}. CM's model predicts the coexistence of an inner adhesive circular region of radius $c$, which decreases with $Q$ according to 

\begin{equation}
c = a_H \left( 1-\frac{Q}{\mu P} \right)^{1/3},
\label{cqCM}
\end{equation}
surrounded by an outer slip annulus, which is in full qualitative agreement with our experimental results. Using a superposition principle, CM's calculations provide complete analytic expressions for $\sigma_{xz}$, $\sigma_{yz}$, $u_x$ and $u_y$ within the contact, in both stick and slip regions \cite{JohnsonBook,HillsBook}. Further derivations by Johnson \cite{JohnsonBook} also give $u_x$ and $u_y$ outside the contact, thus providing the entire displacement field at the interface. Note that CM's model has previously been supported by global force and displacement measurements as well as by wear trace inspection \cite{Johnson1955,Johnson1961}. However, no comparison had yet been performed on the displacement and stress distributions at mesoscopic scales.

\begin{figure}
\resizebox{1.0\hsize}{!}{\includegraphics*{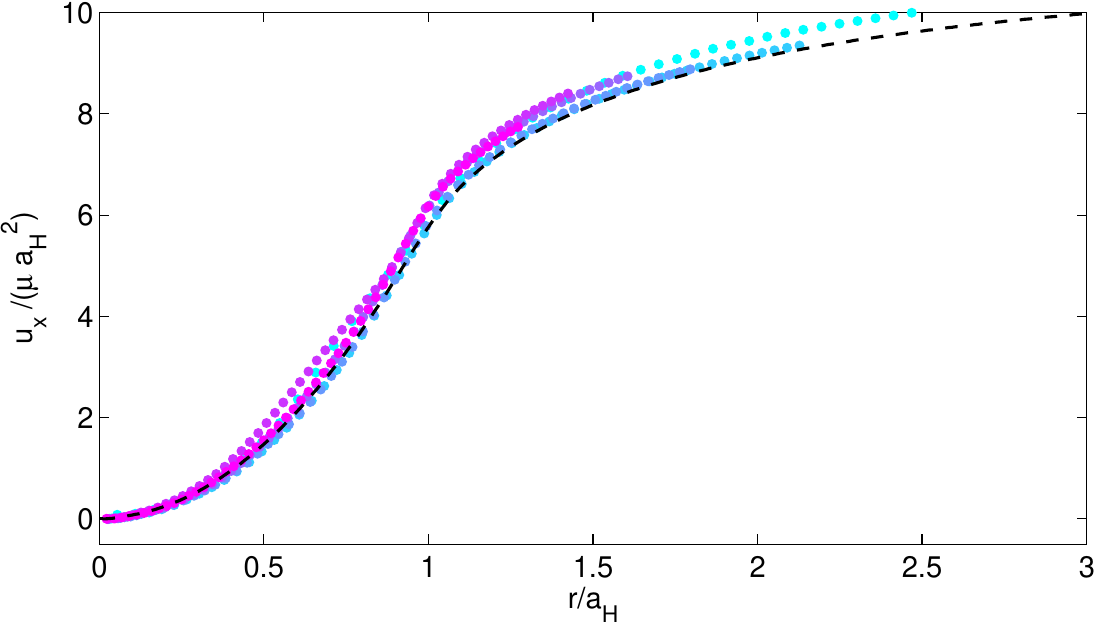}}
\caption{\label{Fig11} (Color online) $u_x/(\mu {a_H}^2)$ in steady sliding versus $r/a_H$ at all $P$ (the color code is the same as in Figs.~\ref{Fig3} and \ref{Fig5}). The dashed black line represents Johnson's prediction with $E=3.43$~MPa, smoothed out with the correlation box size $\lambda$. The value of $\mu$ is obtained by averaging over all images in steady sliding, \textit{i.e.} for all $t \geq t_s$.}
\end{figure}

\begin{figure}
\resizebox{1.0\hsize}{!}{\includegraphics*{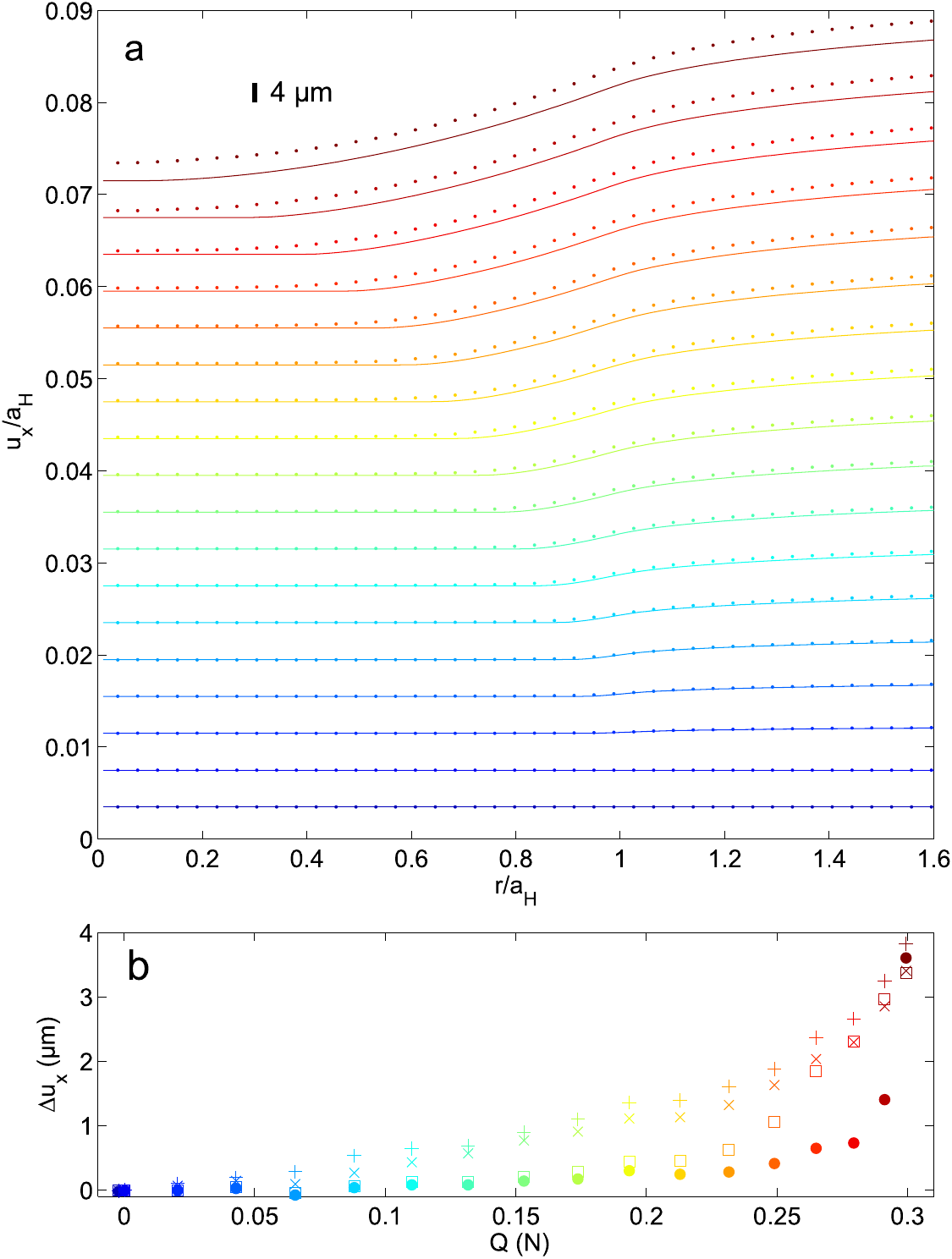}}
\caption{\label{Fig12} (Color online) (a) $u_x/a_H$ versus $r/a_H$ during the loading phase for $P=0.3$~N ($a_H \approx 1.857$~mm), shown for one out of 8 images, \textit{i.e.} every 1 second. Points are the measured displacements. Solid lines represents Cattaneo-Mindlin's model predictions with $E=3.43$~MPa. All curves have been arbitrarily shifted vertically to ease visualization. (b) Deviation ${\Delta}u_x$ between $u_x(r)$ measured and CM's predictions versus $Q$ for $r \sim 0~(\bullet)$, $r=0.5 a_H~(\Box)$, $r=a_H~(\times)$, and $r=1.5 a_H~(+)$. The color code is the same as in (a).}
\end{figure}

Figures~\ref{Fig11} and \ref{Fig12} show a direct comparison (\textit{i.e.} without any adjustable parameter) between the measured and predicted displacement fields, averaged over the azimuthal angle $\theta$, in steady sliding and during the loading phase, respectively. In steady sliding (Fig.~\ref{Fig11}), all $u_x(r)$ curves at all loads have been rescaled by $\mu {a_H}^2$, with $\mu=<\mu_d>=<Q(t \ge t_s)>/P$ and $a_H$ is the Hertz contact radius computed using the value of the Young's modulus $E$ deduced from the JKR test, \textit{i.e.} $E=3.43$~MPa. The agreement with CM is found good at all normal loads $P$. During the transient loading (Fig.~\ref{Fig12}), a similar overall agreement is achieved. However, a closer look reveals that significant deviations occur as $Q$ increases, becoming more pronounced as $Q$ reaches $Q_s$. To quantify these deviations, Fig.~\ref{Fig12}b displays, on the example of $P=0.3$~N, the difference ${\Delta}u_x(r)$ between the measured displacements and CM's predictions at 4 different positions, 3 within the contact ($r = \left\lbrace0, 0.5a_H, a_H\right\rbrace$) and 1 outside of it at $r = 1.5a_H$, as a function of $Q$. As clearly shown, ${\Delta}u_x(r)$ increases with $Q$ reaching a maximal value of a few $\mu$m when $Q \sim Q_s$ at all points $r$. Such deviations will be extensively discussed further down.

\begin{figure}
\resizebox{1.0\hsize}{!}{\includegraphics*{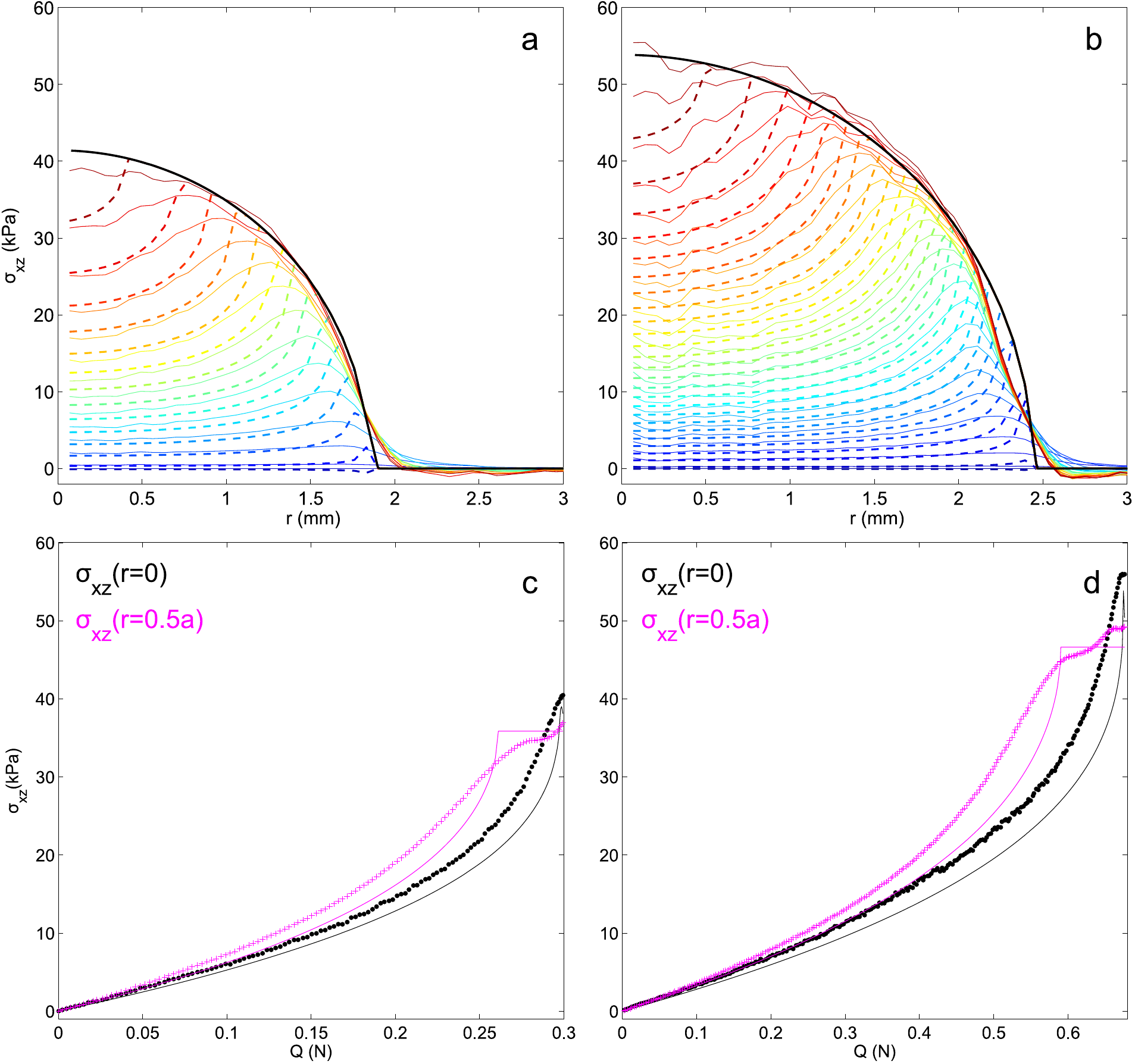}}
\caption{\label{Fig13} (Color online) Angularly averaged stress fields $\sigma_{xz}(x,y)$ at $P=0.2$~N (a) and $P=0.7$~N (b). On both plots, the dashed line curve represents $\mu_s \sigma_{zz}(r)$ where $\sigma_{zz}(r)$ is here the Hertz pressure profile taking for $E$ the optimum values obtained from the inversion procedure and given in the inset of Fig. \ref{Fig9}. Small $Q$ curves are shown in blue, while high $Q$ curves are in red. Profiles are shown every 1.25 second. CM's stress profiles were smoothed out with the correlation box size $\lambda$. (c), (d) $\sigma_{xz}(r=0)$ and $\sigma_{xz}(r=0.5a)$ versus $Q$ for the data shown in (a) and (b) respectively.}
\end{figure}

\begin{figure}
\resizebox{1.0\hsize}{!}{\includegraphics*{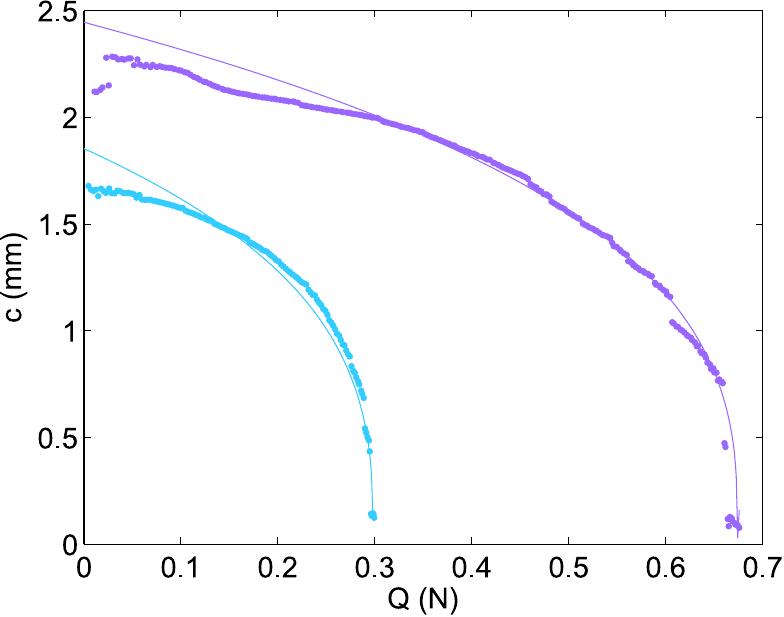}}
\caption{\label{Fig14} (Color online) Tangential load dependence of $c~(\bullet)$ compared to CM's model predictions (thin solid lines, Eq. \ref{cqCM}) for two normal loads, respectively $P=0.2$~N and $P=0.7$~N. The color code is the same as used in Figs.~\ref{Fig3} and \ref{Fig5}. Sub-pixel determination of $c$ was obtained by fitting $\sigma_{xz}$ around its maximum value with a second order polynomial.}
\end{figure}

Similarly, the tangential stress fields $\sigma_{xz}(x,y)$ were angularly averaged to obtain $\sigma_{xz}(r)$ at all shear loads $Q$ (Fig.~\ref{Fig13}). When $Q = Q_s$, if one assumes that Amontons-Coulomb's friction law remains valid at a local length scale, one expects the shear and normal stresses to be related with $\sigma_{xz} = \mu_s \sigma_{zz}$. Taking Hertz's pressure profile\footnote{In Hertz's contact theory, $\sigma_{zz}(r)=\sigma_0 \sqrt{1-r^2/{a_H}^2}$ with $\sigma_0=\left(\frac{6 P E^{*2}}{\pi^3 R^2}\right)^{\frac{1}{3}}$ and $E^*=\frac{E}{1-\nu^2}$ is the reduced modulus.} for $\sigma_{zz}$, we have computed $\mu_s \sigma_{zz}$. Figures~\ref{Fig13}a and \ref{Fig13}b clearly show a rather good agreement between Hertz (black solid line) and our experimental measurements, except for a small tail at the edge of the contact. The latter presumably results from the multi-contact characteristics of the interface, and is to be related to the tail in the intensity profiles as discussed earlier in section \ref{subsubsec:3:1}. When $Q < Q_s$, the stress profiles have qualitatively the same radial dependence as predicted by CM's model and indicated with the dashed lines (Fig.~\ref{Fig13}a and b). Quantitative analysis however reveals that deviations are clearly present as shown for $\sigma_{xz}(r=\{0,0.5a\})$ on Figs.~\ref{Fig13}c and d. Yet, these stress profiles provide a mean to extract the diameter of the stick region $c$. In CM's model, $\sigma_{xz}$ is maximum at $r=c$. Assuming that it is still true in our case, it is thus possible to give an estimate of $c$ and compare it to CM's predictions given by Eq. \ref{cqCM} (Fig.~\ref{Fig14}). Clearly, the agreement for such a macroscopic quantity is good, even though the amplitudes of $\sigma_{xz}$ are not exactly captured by CM's stress predictions.   

\section{Local constitutive law of friction}
\label{sec:5}

At this point, we have shown that our measurements agree with CM's model, not only qualitatively with the existence of an inner stick region surrounded by a growing annulus of slip, but also quantitatively since both displacement and stress distributions are found to follow reasonably well the predicted shape and amplitude. However, close comparison reveals discrepancies, the most striking being that the displacement in the vicinity of the center of the contact is not strictly null during the transient phase, but slowly increases with $Q$ as evidenced in Fig.~\ref{Fig12}b. This observation is at odds with the rigid-plastic-like constitutive law assumed in CM's model, which implies that the displacement $u_x$ at the center of the contact should remain null as long as $\sigma_{xz} \leq \mu_s \sigma_{zz}$. A possible explanation can be found when considering the principle of the DIC measurement itself, which is based on correlation boxes containing pixels corresponding to both micro-contacts and out-of-contact regions. Contrary to experiments with smooth marked PDMS such as in \cite{Chateauminois2008,Chateauminois2010}, the measured displacements are thus averaged over the thickness $h$ of the rough layer and combine two intricate contributions: true slip at the micro-contacts and elastic shear deformation of the rough layer.

\begin{figure}
\resizebox{1.0\hsize}{!}{\includegraphics*{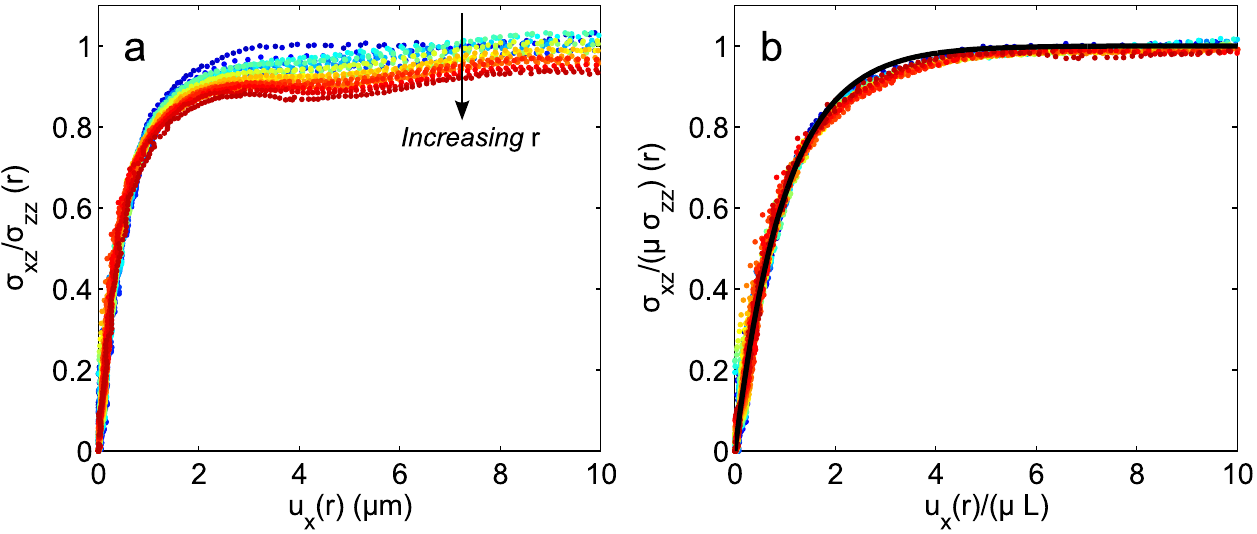}}
\caption{\label{Fig15} (Color online) (a) $\frac{\sigma_{xz}}{\sigma_{zz}}$ versus $u_x$ for $P= 0.7$~N and at different $r$ taken every 70 $\mu$m between 0 and $a_H - \sqrt{Rh}$. Small $r$ valued curves are shown in dark blue, while largest $r$ valued curves are in dark red. (b) $\frac{\sigma_{xz}}{\mu \sigma_{zz}}$ versus $u_x/(\mu L)$ for the same load at different $r$. The solid black curve is $1-e^{-\alpha}$ with $\alpha$ a dimensionless increasing number and corresponds to the predictions of Eq. \ref{EqnMacro}.}
\end{figure}

In order to characterize the local shear behavior of the interface, we examine the relationship between the tangential stress $\sigma_{xz}$ and displacement $u_x$ measured at various positions $r$ within the contact. Assuming that the pressure profile $\sigma_{zz}$ is given by Hertz's contact theory, we compute the ratio $\frac{\sigma_{xz}(r)}{\sigma_{zz}(r)}$ as a function of the displacement at positions $r$, $u_x(r)$. Figure~\ref{Fig15}a shows, on the example of $P=0.7$N, the typical measured behavior obtained for all normal loads $P$. Two distinct regimes can be identified. At small $u_x$, $\sigma_{xz}/\sigma_{zz}$ increases quasi-linearly with $u_x$. At large $u_x$ however, $\sigma_{xz}/\sigma_{zz}$ asymptotically reaches a constant value. This observation can be understood as a direct consequence of the shear response of the multi-contact interface. This problem has been theoretically analyzed by Bureau and coworkers in the context of a plane-on-plane frictional contact configuration \cite{BureauPRAS2003}. The surface topography is described using Greenwood-Williamson's model \cite{GreenwoodWilliamson1966} by an assembly of spherical asperities of equal radius, and whose heights are randomly distributed with an exponential distribution. In addition, the response of each asperity upon tangential loading is described by CM's model. The macroscopic normal and shear forces $P$ and $Q$ are distributed uniformly over the multi-contact interface. The model predicts the evolution of the ratio $Q/P$ as a function of the relative displacement $\delta$ of the centers of mass of both solids in contact, in the form:

\begin{equation}
\frac{Q}{P}=\mu \left( 1-e^{-\frac{\delta}{\mu L}} \right)
\label{EqnMacro} 
\end{equation}
where $\mu$ is a microscopic friction coefficient and $L=\frac{2-\nu}{2(1-\nu)}h$ is an elastic length whose value is controlled by the \textit{rms} roughness of the interface $h$, and which depends on the material properties only through the Poisson's ratio $\nu$. This prediction allowed the authors to interpret global friction force \textit{vs} displacement measurements at the interface between two Plexiglas surfaces submitted to minute shear oscillations.

In our sphere-on-plane configuration, the interfacial stress field is not uniform. However, one may expect that at intermediate length scales, smaller than the macroscopic contact size but larger than the inter-asperity distance, the local ratio between both components $\sigma_{xz}$ and $\sigma_{zz}$ also obeys Eq. (\ref{EqnMacro}). The validity of this result is demonstrated in Fig.~\ref{Fig15}, in which the local stress ratio is plotted as a function of the measured displacement, for various radii $r$ and for $P=0.7$~N (Fig.~\ref{Fig15}a). The different curves exhibit very similar behaviors. However a slight pressure dependence is visible at large displacements, which can be corrected for by dividing each curve using the asymptotic friction coefficient $\mu$ (\textit{i.e.} the shear-to-normal stress ratio measured at large $u_x$). This yields a master curve, shown in Fig.~\ref{Fig15}b, which can then be fitted using Eq. \ref{EqnMacro}. Combining all 6 experiments at different normal loads $P$, the fitting parameters are found to have very little dependence with $\sigma_{zz}$ with $\mu = 0.93 \pm 0.08$ and $L = 0.80 \pm 0.23~\mu$m. The local friction coefficient $\mu$ is found consistent with the macroscopic static friction coefficient $\mu_s$ (\textit{see} Fig.~\ref{Fig8}). The value of the elastic length is to be compared with $L=0.89~\mu$m predicted within Bureau \textit{et al.}'s model when considering the Poisson's ratio $\nu$=0.5 of PDMS and the $rms$ roughness $h=0.595~\mu$m of our surface (\textit{see} Section \ref{subsec:2}).

This quantitative agreement suggests that Bureau \textit{et al.}'s macroscopic model for the shear response of multicontact interfaces can be extended to mesoscopic scales. It also shows that the observed deviations from CM's model are fully compatible with an elasto-plastic-like behavior of the rough interface, not taken into account in Amontons-Coulomb friction law. In this respect, the type of measurements performed here can be used to estimate the shear stiffness $k$ of a multicontact interface, a quantity which has recently received significant attention (see \textit{e.g.} \cite{GonzalesValades2010,Akarapu2011,Campana2011,Kartal2011}). By extending Bureau \textit{et al.}'s model locally, one can define $k$ as the initial slope of the curves in Fig.~\ref{Fig15}, \textit{i.e.} $k=\left(\frac{\partial \sigma_{xz}}{\partial u_x}\right)_{u_x=0}=\frac{\sigma_{zz}}{L}$. Taking 25kPa as a typical pressure value in our experiments, one finds $k \sim$ 30 kPa/$\mu$m to be the corresponding typical shear stiffness.

A few additional experiments were performed for a different roughness ($h \sim 1.3~\mu$m) and different driving velocities up to $20~\mu$m/s in order to test the robustness of these measurements. In the transient regime ($Q \le Q_s$), the results appear to be consistent with the observations reported in this work (Figs.~\ref{Fig11}--\ref{Fig14} and Fig.~\ref{Fig15}), provided the data is rescaled with respect to the velocity and maximum shear force. In contrast, the long time behaviors ($Q > Q_s$) appears to depend on both the driving velocity and the elastomer's curing protocol \cite{Kurian2010}. The characteristics of these long transients remain unexplained to date and are beyond the scope of the present work.    

\section{Conclusion}
\label{sec:6}

Building on Chateauminois \textit{et al.}'s earlier work \cite{Chateauminois2008,Chateauminois2010}, we have developed a novel imaging technique to probe locally the spatial distribution of tangential displacement and associated shear stress averaged over the micrometric thickness of a heterogeneous \textit{multi-contact} interface. The method, based on Digital Image Correlation, uses both micro-contacts and micro-asperities as displacement tracers, yielding a sub-micrometer resolution in the measured displacement with a typical 140 $\mu$m spatial resolution. We emphasize that this technique is only suited for rough elastic surfaces against smooth rigid bodies.

This method was used to study the transition from static friction to macroscopic sliding of a smooth glass sphere tangentially loaded against a microscopically rough elastomer plane. This model geometry allowed us to directly compare the measured micro-mechanical fields to a classical model by Cattaneo and Mindlin. We showed that their model does capture reasonably well both the shape and amplitude of the measured displacement and stress fields. However, close comparison reveals that significant deviations occur, which have been shown to involve a characteristic length scale of the order of the micrometric surface roughness. In this respect, we characterized the elastic shear response of our multi-contact interface prior to slippage, which is ignored in CM's model. The latter was shown to be well captured by the model of Bureau \textit{et al.}'s developed in \cite{BureauPRAS2003}.

Overall, the present study suggests the need to replace the rigid-plastic-like Amontons-Coulomb friction law with an elasto-plastic constitutive friction law in CM-like derivations of the displacements/stress fields, and more generally in any micromechanical analysis of contact mechanics problems (as done in \textit{e.g.} \cite{Brzoza2008}). The effective modulus of the elastic part of this constitutive law is (i) proportional to the local applied pressure and (ii) inversely proportional to the thickness of the rough interfacial layer. The type of measurements developed and validated in this work opens the way for more focused studies in any other contact geometry or loading configurations, for which no explicit model might be available. The time-resolution of the measurements being entirely controlled by the frame-rate of the imaging system, we anticipate that the very same method could also be used in the fast transient regimes involved in frictional instabilities.

\begin{acknowledgement}
The authors are indebted to Antoine Chateauminois of Laboratory PPMD of ESPCI for his valuable help on the inversion procedure and to both Antoine Chateauminois and Christian Fr\'etigny of PPMD, ESPCI for fruitful discussions. This work was partly supported by ANR-DYNALO contract NT09-499845.
\end{acknowledgement}

\end{document}